\documentclass[twocolumn,prl,floatfix]{revtex4}
\usepackage{amsbsy}
\usepackage{amstext}
\usepackage{amsmath}
\usepackage{graphicx}
\usepackage{color}

\begin{document}

\title{Nemato-elastic crawlers}
\author{A.~P. Zakharov and L.~M. Pismen}
\affiliation{Technion -- Israel Institute of Technology, Haifa 32000, Israel} \email{pismen@technion.ac.il} 

\begin{abstract}
{A propagating ``beam'' triggering a local phase transition in a nematic elastomer sets it into a crawling motion, which may morph due to buckling. We consider the motion of  the various configurations of slender rods and thin stripes with both uniform and splayed nematic order in cross-section, and detect the dependence of the gait and speed on flexural rigidity and substrate friction.}

pacs{ 46.32.+x, 46.70.De, 02.40.Yy, 61.30.Jf}  

\end{abstract}
 
 \maketitle

Liquid crystal elastomers (LCE), made of cross-linked polymeric chains with embedded mesogenic structures combine orientational properties of liquid crystals with shear strength of solids. This biomimetic material was first envisaged by de Gennes \cite{degennes} as a prototype for artificial muscles, and synthesised by Finkelmann and co-workers \cite{Finkelmann}. The specific feature of LCE is a strong coupling between the director orientation and mechanical deformations \cite{Warner}, which can be controlled by the various physical and chemical agents to enable transfer of chemical or optical inputs to mechanical energy. When the material undergoes a phase transition from the isotropic to nematic state, it strongly elongates along the director and, accordingly, shrinks in the normal directions to preserve its volume; the opposite effect takes place as a result of the reverse transition. 

The most common way to induce reversible transition between the nematic and isotropic state (NIT) is photo-isomerisation of the azobenzene moieties between the cis and trans forms \cite{Finkelmann01}. The reshaping induced thereby was explored to produce the various bent forms that may serve as actuators \cite{review10,review12,Yusm12,Ionov}, as well as opto-mechanical transducers producing rotational motion \cite{rotor} and swimming into the dark \cite{swim}. Potential applications extend to biomimetic devices of soft robotics \cite{soft1,soft2,micro}.

Photo-isomerisation is reported to be on the 10 ns time scale \cite{isokinetic}. Changes in the orientational order and mechanical deformations are much slower, and may vary in a wide range. In an optical fiber, fast photomechanical response upon laser irradiation with response times of the order of less than a tenth second was observed \cite{fiber}, while response times in the order of an hour were reported in main-chain LCE \cite{Fink13}. It is not clear whether nematic alignment or mechanical deformation has been a limiting factor in the various experimental setups. Another way of inducing transition is thermal, using LCE nanocomposites containing superparamagnetic nanoparticles that perform local heat dissipation upon irradiation with electromagnetic fields \cite{magnet} on the scale of less than a minute. An apparently unexplored mechanism, applied so far in swelling isotropic gels but not in LCE \cite{Balazs,YT96,KYB_07,YB12}, is mechanical action of oscillating chemical reactions that would generate a wave of  reversible NIT. 

While earlier work concentrated on deformations of monodomain LCE, deformations and stresses in imperfectly ordered materials containing defects have attracted more recent attention. Stresses arising due to these intrinsic deformations were investigated for confined flat sheets where they were shown to lead to phase separation of isotropic and nematic domains \cite{epj13} and formation of persistent defects that are not necessitated by topology \cite{non15}. Internal stresses relax in a natural way when three-dimensional (3D) deformations are allowed. The reshaping effect causes bending of flat thin sheets into curved shells. The simplest symmetric forms emerge in the vicinity of unit charge defects \cite{Warner12,pre14,mw15}, while more complex shapes are generated by naturally occurring half-charged defects \cite{epj15}. Programmed morphing of LCE shells has become possible with the development of  methods of impressing a desired nematic structure in LCE sheets \cite{Broer14,BroerLa,White,WhiteSM} that might make use of solutions of a reverse problem of  computing a desired two-dimensional (2D) metric that would determine a 3D shape upon NIT \cite{sharon}. 

Little is known so far on \emph{dynamic} effects accompanying the transmission of optical or chemical energy into motion in LCE. In this Letter, we study how a propagating wave inducing reversible NIT sets into motion LCE strings or sheets resting on a solid substrate, thereby enabling biomimetic crawlers or walkers. 

\begin{figure}[b]
\begin{tabular}{cc}(a)& (b)\\
\includegraphics[width=.21\textwidth]{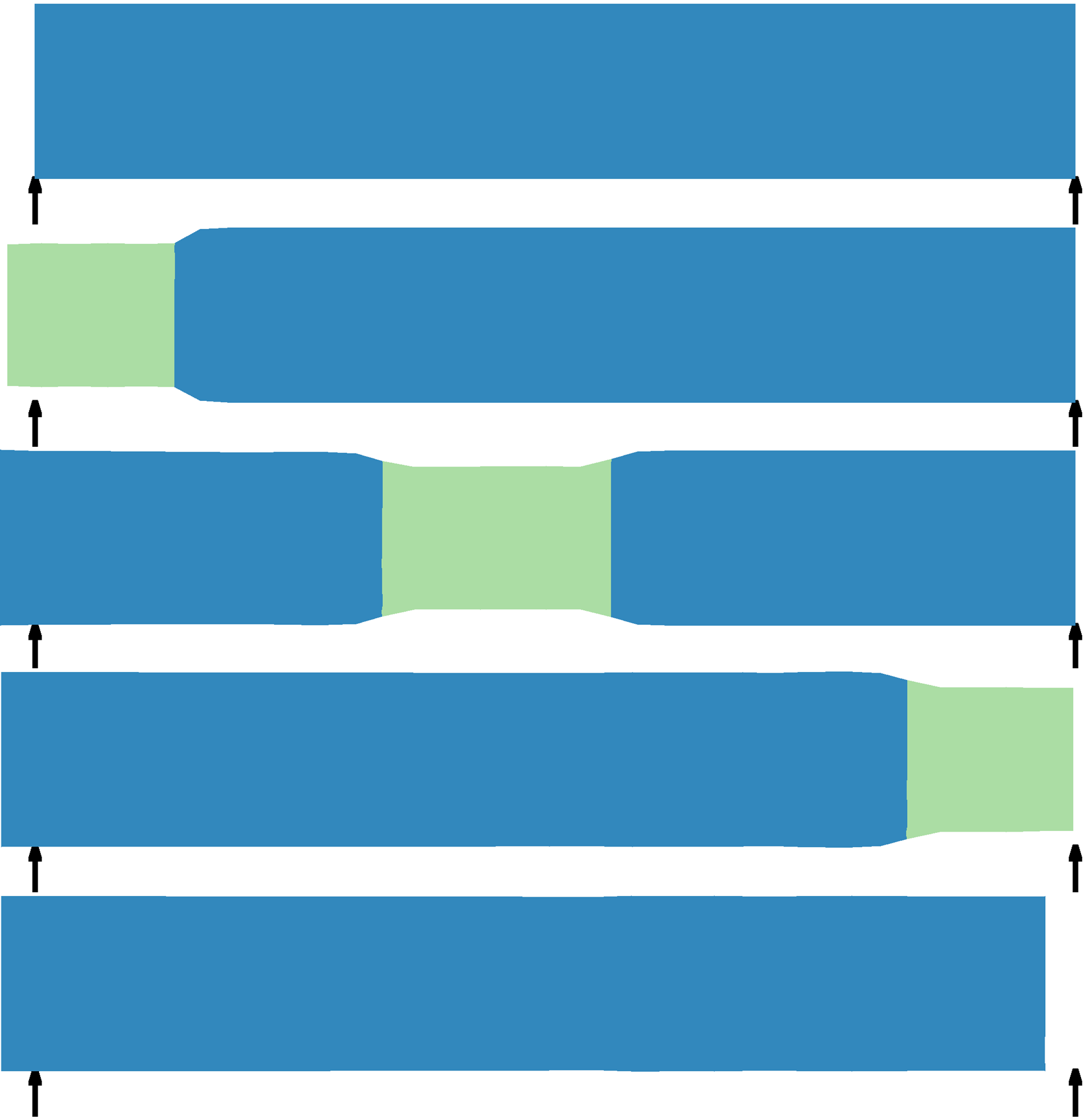} 
& \includegraphics[width=.25\textwidth]{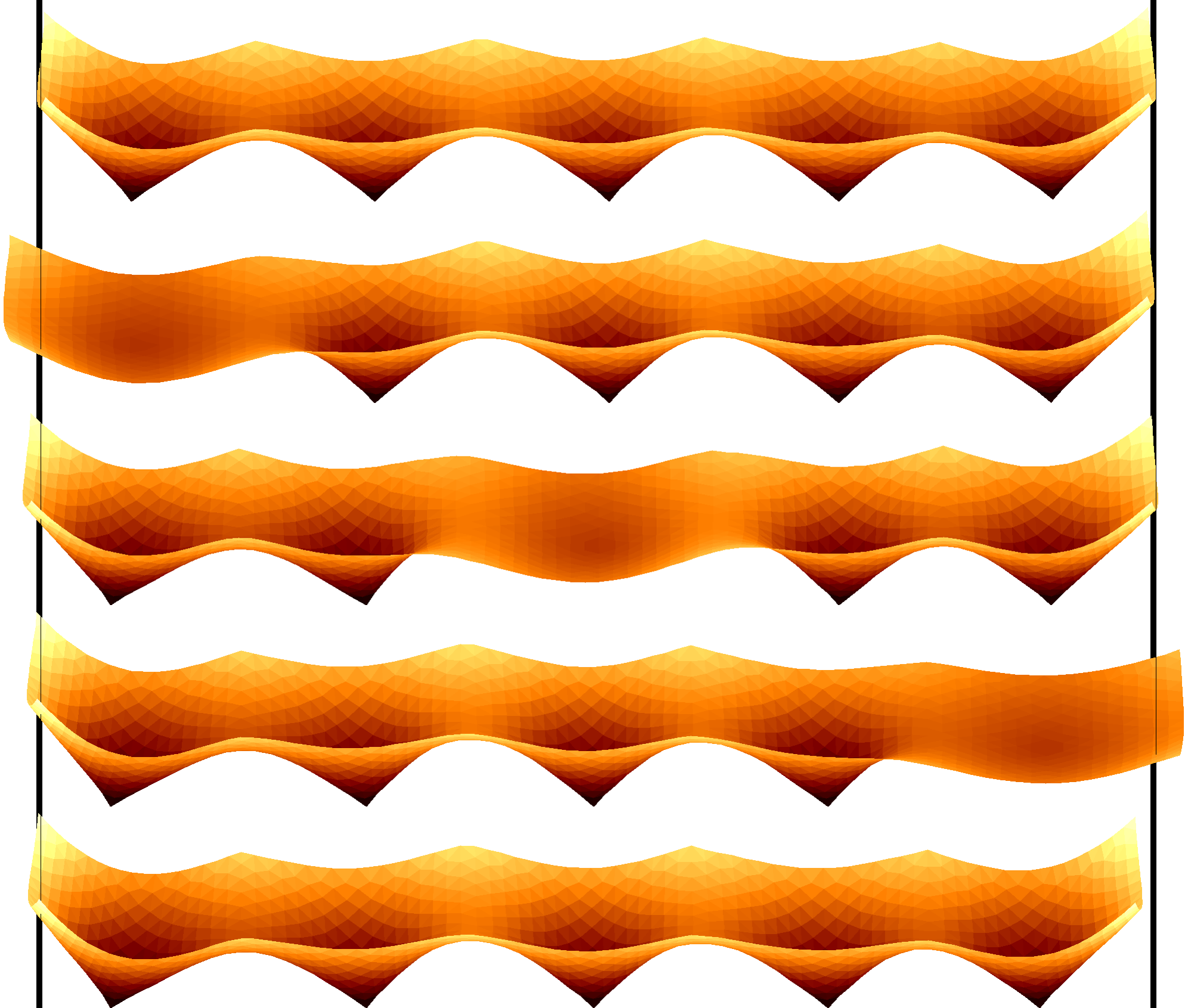}
\end{tabular}
\caption{Successive configurations of an inflexible rod or a hollow cylinder (a) and of  a   walker with five sequentially actuated conical legs (b) on a substrate with a very high friction under a beam propagating from the left. Vertical lines mark the original positions of the ends.}
\label{f:worm}
\end{figure}

The simplest case, illustrating how an LCE ``worm" may crawl, is an inflexible rod, initially in the isotropic state, resting on a substrate with a very high friction. We apply a dynamic \emph{localized} perturbation, further referred to as a ``\emph{beam}", enforcing a local phase transition. This notion is understood in a generalized sense, which may imply a thermal or chemical, as well as optical excitation. As a beam of the length $L$ comes from the left, triggering the transition (assumed, for simplicity, to be instantaneous and uniform in cross-section) to the nematic state polarized along the rod, the ``illuminated" part extends by a factor $\lambda$ and narrows, thereby losing the contact with the substrate. As the beam passes, the left end reverts to the isotropic state, thickens back and becomes immobile, while the narrowing propagates to the right and eventually exits at the right end leaving the worm translated against the beam propagation direction by the distance $(\lambda-1) L$ (Fig.~\ref{f:worm}a). If the frequency of the beam passage is $\omega$, the velocity of the rod is  $\omega(\lambda-1) L$. A rectangular stripe moves exactly in the same way, provided the beam propagates along its axis and covers its width uniformly, and so does a peristaltically deforming hollow cylinder with a uniform nematic state around the circumference, the only substantial difference being a gradual change of the girth at the beam edges depending on the flexibility of the shell. 

The same considerations apply qualitatively to more complex (and computationally more difficult) structures created by programmed morphing, such as a ``voxellated" walker formed by imprinting a vortex texture with a unit-charge defect in an assembly of  square units \cite{White}. Here the starting state is the conical structure obtained after NIT, while the ``beam"  enforces a reverse transition subsequently in each square.  Similar to a segment of an inflexible rod, a square under the beam extends, and the voxellated walker moves by lifting its foot under the beam (Fig.~\ref{f:worm}b). The position is shifted against the beam propagation direction, since a shift in the opposite direction is counteracted by the friction and flexural rigidity of  the next conical unit in line. Shortening under the beam acts in the opposite way, causing motion along the beam propagation. Crawling motion, though slow, is very stable, compared to wheeled and legged machines. Crawling robots have a low load concentration per unit area, and elongation combined with a decreasing thickness enhance their ability to penetrate narrow interstices. A larger variety of gaits is possible due to spontaneous bending or asymmetric transitions in two-dimensional (2D) sheets, as will be detailed below. 

Our modeling technique  mainly follows that employed in studies of relaxation of LCE to equilibrium \cite{epj15,ZSP_PRE_11,korea15}, based on Lagrangian finite element computation with overdamped dynamics minimizing an assigned energy functional. Unless stated otherwise, 2D and 1D computations start, respectively, from a flat thin sheet or a straight slender rod, taken as the reference state. The elastic energy density of a sheet or rod deformed following NIT (scaled by the shear modulus $\mu$) is 
\begin{equation}
\mathcal{E} = \frac 12  (u_{ij}-\overline{u}_{ij})^2 
 +  \frac {1}{2c}  h^2 (\kappa_{ij}-\overline{\kappa}_{ij})^2 + G z.
 \label{eq:Elast1}
\end{equation}
Here $u_{ij}=\frac 12 (x_{i,j}+x_{j,i})-\delta_{ij}$  is the deformation tensor, $x_{i,j}$ is the strain tensor, defined as the partial derivative (denoted by a comma) of the current position $x_i$ with respect to the reference coordinate $X_j$,  $\delta_{ij}$ is the Kronecker delta; $\kappa_{ij}$ is the curvature tensor,  $\overline{u}_{ij}$ and $\overline{\kappa}_{ij}$ are the intrinsic deformation and spontaneous curvature due to a phase transition, $h$ is the local shell thickness or rod radius, and $c$ is a numerical factor equal to 9 for a sheet and 16 for a cylindrical rod with, respectively, the thickness or diameter $h$ \cite{LL}. All lengths will be scaled by the beam width $L$, which has to far exceed the thickness of a rod or sheet; therefore excess bending is strongly favoured over in-plane elastic deformation. The last term expresses the scaled gravitational energy, where $G=g\rho L/\mu$, $g$ is the acceleration of gravity, $\rho$ is density, and $z$ is the elevation over the substrate plane. This term is likewise small compared to in-plane elastic energy but may be comparable to the flexural energy. We will neglect the energy due to a gradient of the nematic order parameter, which arises in the transitional zones at the edges of the beam. An additional nematic gradient energy induced by curvature is commonly negligible compared with the elastic bending energy. 

Similar to Ref.~\cite{epj15}, the overall scaled energy $\mathcal{F}=\int h \mathcal{E} d^2\mathbf{X}$ is discretised in 2D on a regular triangular grid in the reference state with the nematic director field defined on the nodes. The in-shell deformation energy (scaled by $\mu L^3$) given by the first term in Eq.~\eqref{eq:Elast1} is presented then as
\begin{equation}
\mathcal{F}_s = \frac {1}{2} \sum_n^\mathrm{edges}{ h_n} (l_n -\overline{ l}_n)^2,
 \label{eq:ElastEn}
\end{equation}
where $l_n$ and $\overline{ l}_n$ are, respectively, the observed length of an $n$th edge and its ``optimal" length due to intrinsic elongation following NIT. If the length of an edge parallel to the director shortens by the factor $\lambda>1$, the edge normal to the director, as well as the shell thickness, elongate by the factor $\lambda^{1/2}$ to preserve the volume. Accordingly, the value $\overline{ l}(\psi)$ for an edge of unit reference length at an angle $\psi$ to the director is given by 
\begin{equation}
\overline{ l}^2 =\frac 14\left(\sqrt{\lambda }-\frac{1}{\lambda}\right)^2  \sin^2 2\psi + 
 \left(\sqrt{\lambda }\sin ^2 \psi+\frac{\cos^2 \psi}{\lambda} \right) ^2 .
\label{eq:normmt}
\end{equation}
The curvature tensor at a node with a vertical coordinate $z=f(\mathbf{x})$ is computed as a discretization of the general formula $\kappa_{ij} = f_{,ij}/(1+ f_{,k}^2)^{3/2}$. In 1D computations, assuming polarization along the axis of the rod, the shortening along the axis upon NIT equals to $\lambda^{-1}$, and Eq.~\eqref{eq:ElastEn} with $\overline{ l}_n= l_n/\lambda$ and $h_n$ replaced by $h_n^2$ is applicable.

General dissipative dynamic equations obtained by varying the overall energy with respect to the node positions have the form 
\begin{equation}
 \frac{d \mathbf{x}_n}{dt } = - \overline{\eta}\frac{\delta\mathcal{F}}{\delta \mathbf{x}_n}, %
 \label{eq:fric}\end{equation}
where the mobility $\overline{\eta}$ is reduced from the reference value $\overline{\eta}=1$ to $\eta<1$ when a node resting on the substrate is displaced horizontally. 
The evolution was followed using the method of gradient descent, whereby new node coordinates were computed by numerical variation of nodes positions. It was assumed that both NIT and the reverse transition are fast. At each small displacement of a beam effecting NIT, deformations were relaxed to their equilibrium values. NIT was assumed to be reversible, so that the nematic order returns to its original state after the beam has passed.

\begin{figure}[t]
\begin{tabular}{cc} (a)& (b)\\
\includegraphics[width=.27\textwidth]{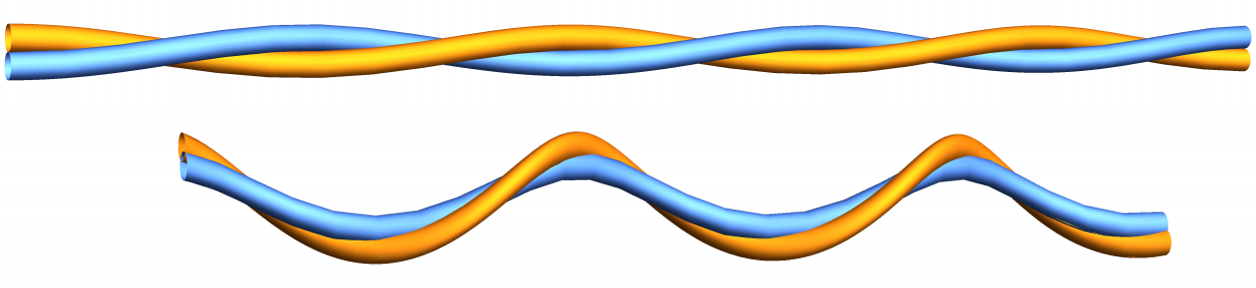} 
& \includegraphics[width=.2\textwidth]{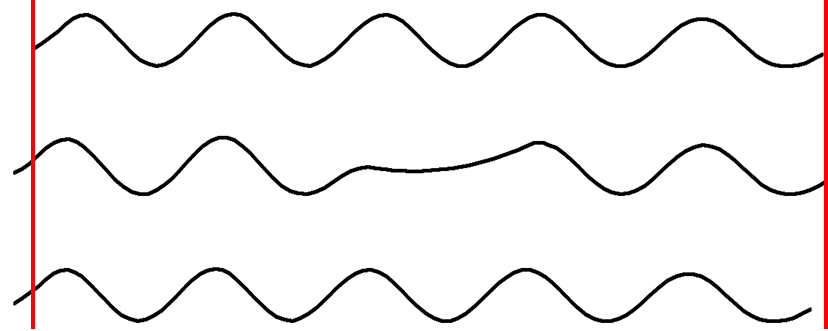} 
\end{tabular}
\caption{(a) A twisted yarn in the isotropic state (above) and after nematic transition (below). (b) Snapshots of the contact line of a twisted yarn on a substrate with a very high friction under a beam propagating from the left, starting from the nematic state (side view). Vertical lines mark the original positions of the ends.}
\label{f:fiber}
\end{figure}

\begin{figure}[b]
\begin{tabular}{ccc}(a)& (b)& (c)\\
\includegraphics[width=.18\textwidth]{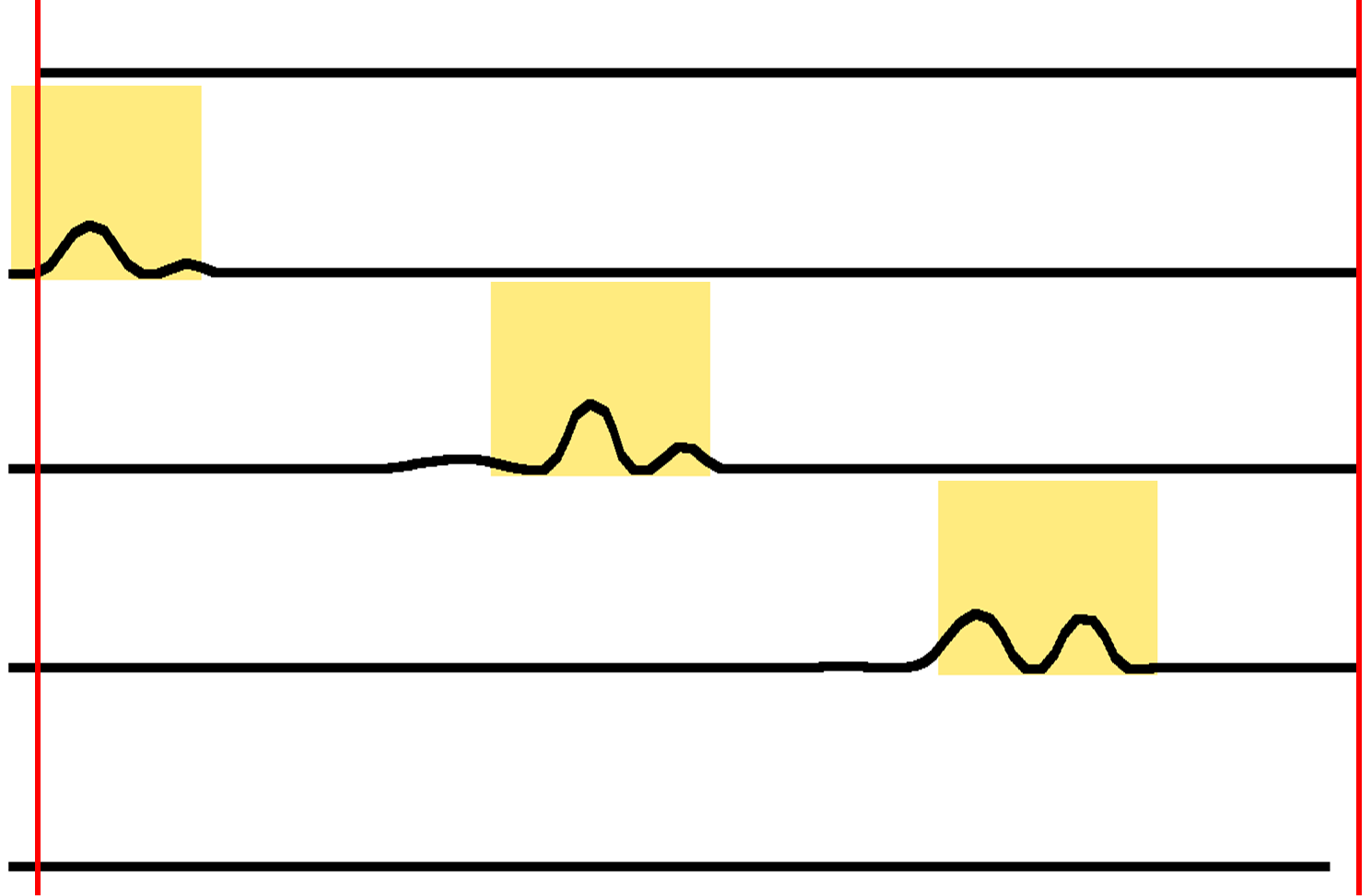} 
& \includegraphics[width=.085\textwidth]{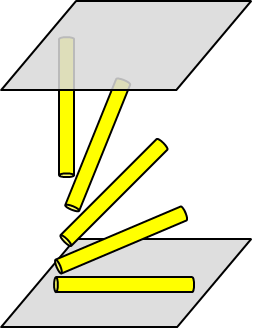}
& \includegraphics[width=.18\textwidth]{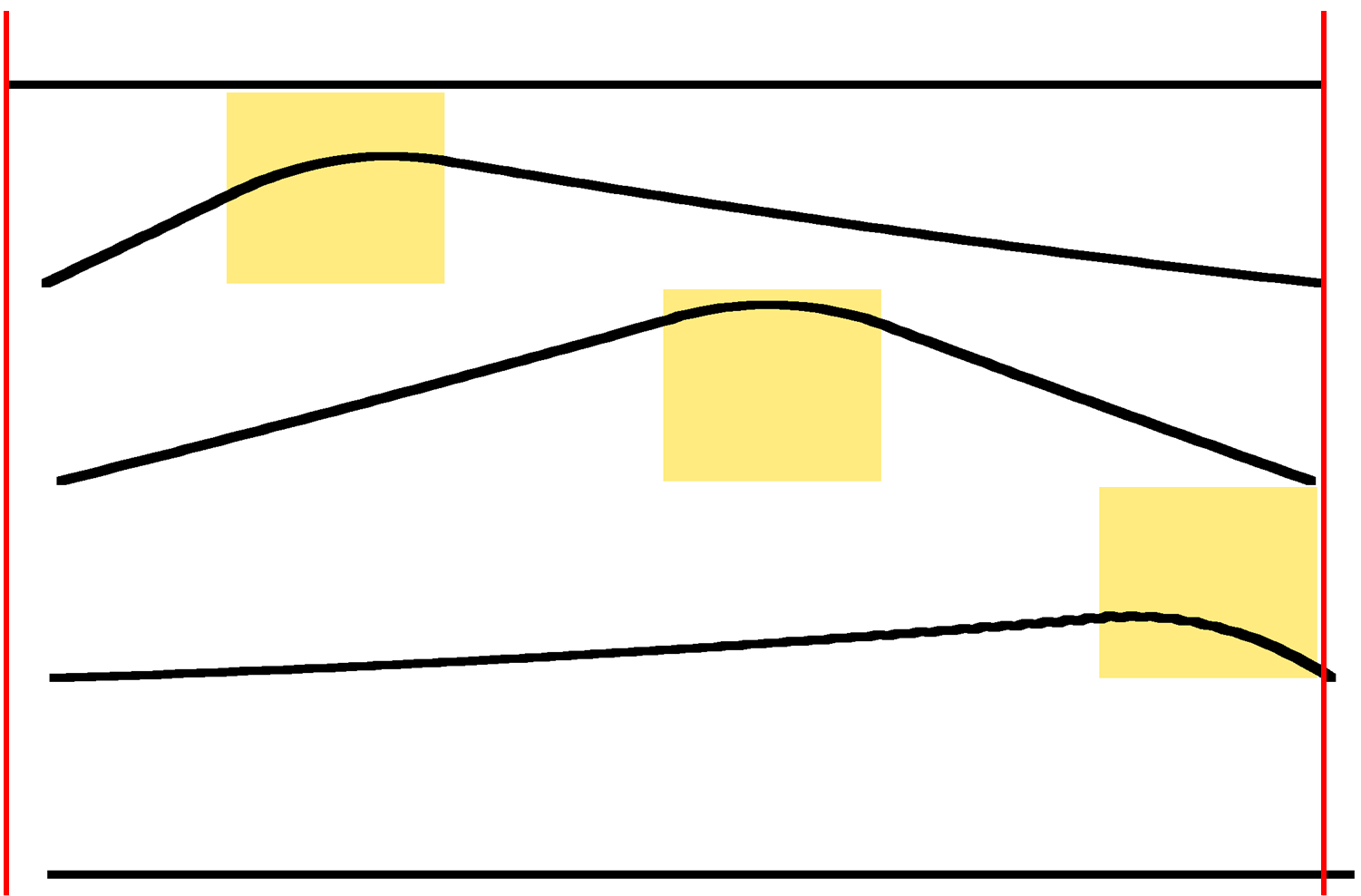}\\
\end{tabular}
\begin{tabular}{cc}(d)& (e)\\
\includegraphics[width=.24\textwidth]{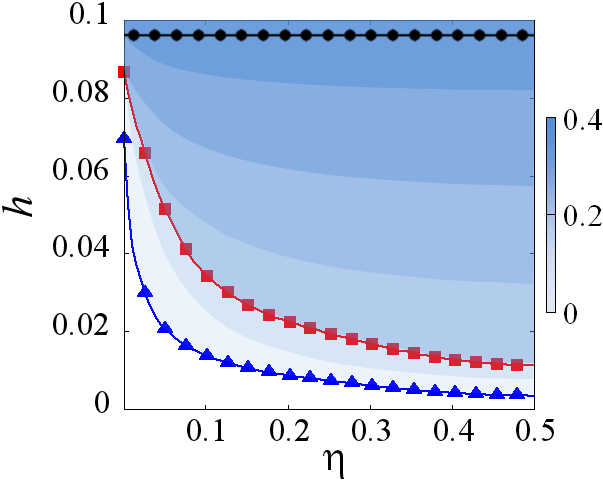} 
& \includegraphics[width=.24\textwidth]{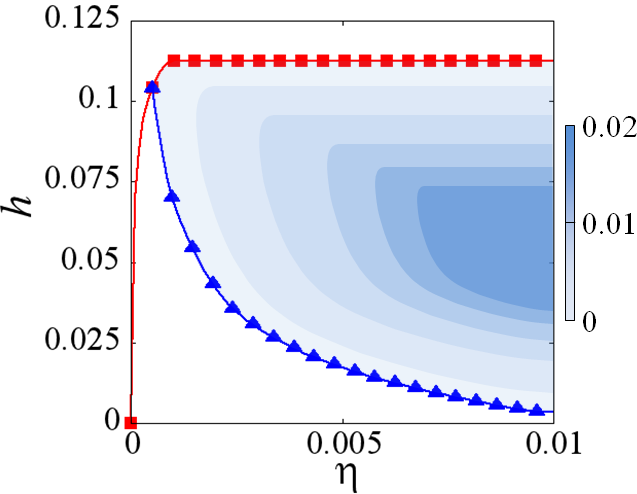}
\end{tabular}
\caption{Above: successive shapes and positions of  a translating flexible rod with uniform (a) and splayed (c) transverse director orientation. Vertical lines mark the original position of the ends. Shaded regions mark instantaneous beam positions. (b) Cross-sectional director orientation in a splayed nematic. Below: bifurcation lines of translational and bending transitions (marked, respectively, by triangles and squares) of a flexible sheet with uniform (d) and splayed (e) transverse director orientation. Shading shows the values of translational velocity. Parameters: (d) $\lambda=1.4$, $\eta=0.2$, $G=0.01$;  (e) $\lambda=1.01$, $\eta=0.1$, $G=0.01$.}
\label{f-string}
\end{figure}

Bending necessarily arises upon NIT in the case of a cross-sectionally non-uniform order, which may be caused by induced warping due to incompatible anchoring at the opposite surfaces \cite{Broer14}, or graded light penetration \cite{WarnerMaha03,Warnerbend07}, or use of layered sheets \cite{laminate15,Timoshenko}. Any cross gradient of the nematic order causes upon phase transition spontaneous curvature $\overline{\kappa}_{ij}$, entering the second term in Eq.~\eqref{eq:Elast1}, and even small extension $\lambda-1 \ll 1$ causes strong 3D deformations. 

The simplest example of a non-uniform order is a yarn made of two intertwined fibers with a diameter $h$, one nematic and the other one an isotropic elastomer with identical properties, glued along a contact line (Fig.~\ref{f:fiber}a), which is straight when the nematic finer is in the isotropic state. When this fiber elongates upon phase transition, the curvature develops along the normal vector directed along the line connecting the centers of both fibers, which rotates with the original pitch. As a result, the contact line turns into a helix with the curvature radius verifying the equation $\lambda (R-\frac 12 h)=R+\frac 12 h(1- \lambda^{-1/2})$, or $R \approx  h/\epsilon $ at  $\lambda-1 \equiv \epsilon \lesssim h \ll 1$. Fig.~\ref{f:fiber}b shows successive positions of the yarn, originally coiled in the nematic state under a passing beam locally converting it to the isotropic state. The yarn moves against the beam propagation essentially in the same way as a rod, a cylinder, or a ``voxellated" walker in  Fig.~\ref{f:worm}.

In a splayed nematic with tangential anchoring on the one side and normal on the opposite side of a sheet (Fig.~\ref{f-string}b), the length shortens upon NIT by the factor $\lambda$ along the director on the one side and elongates by $\lambda^{1/2}$ on the opposite side of the sheet. This generates the curvature radius $R$ (at the side with tangential anchoring) verifying the equation $R+\frac 12 h=(R-\frac 12 h) \lambda^{3/2}$, or $R \approx \frac 23 h/\epsilon $ at  $\lambda-1 \equiv \epsilon \ll 1$. Normally to the director, the stripe elongates on both sides, so that the Gaussian curvature remains zero. If, however, the director configuration is twisted with the parallel but mutually normal director orientation on both sides, the two principal curvatures have the opposite sign and the same absolute value, so that the shell acquires the geometry of a saddle with vanishing mean curvature.
 
Spontaneous curvature may also emerge upon NIT when the nematic order is uniform in the cross-section, causing the gait to change as flexural rigidity decreases in a thinner rod or sheet. In the simplest setup with a NIT-inducing beam of unit length propagating along a nematic strip and covering uniformly its entire width, the solution depends only on the longitudinal coordinate $x$. The successive shapes and positions of  a translating flexible uniform or splayed sheet are shown in Fig.~\ref{f-string}a,c. The lower panels of this Figure show the loci of translational and bending transitions depending on the surface mobility $\eta$ and thickness $h$. 

The motion is very much is different in the two cases, with opposite propagation direction. When the director field is uniform across the sheet,  the motion is directed, as for an inflexible rod or sheet, counter the beam propagation. Crawling motion without bending generally proceeds as described qualitatively above but it slows down due to buckling as the flexural rigidity decreases with decreasing thickness, and ceases altogether below the translational transition line in Fig.~\ref{f-string}d. Buckling is suppressed above the bending transition line in the same figure but velocity is still impeded there as the frontal segment bends under gravity sufficiently strongly to touch the substrate. Below the maximal thickness enabling such bending shown in the same Figure, the velocity increases with increasing surface mobility but above this line it remains constant at the achieved maximum value. In the crawling gait, state-dependent friction is essential. Indeed, integrating Eq.~\eqref{eq:fric} with no stress boundary conditions shows that the average displacement vanishes when friction is constant. 

\begin{figure}[t]
\begin{tabular}{cc} (a)& (b)\\
\includegraphics[width=.24\textwidth]{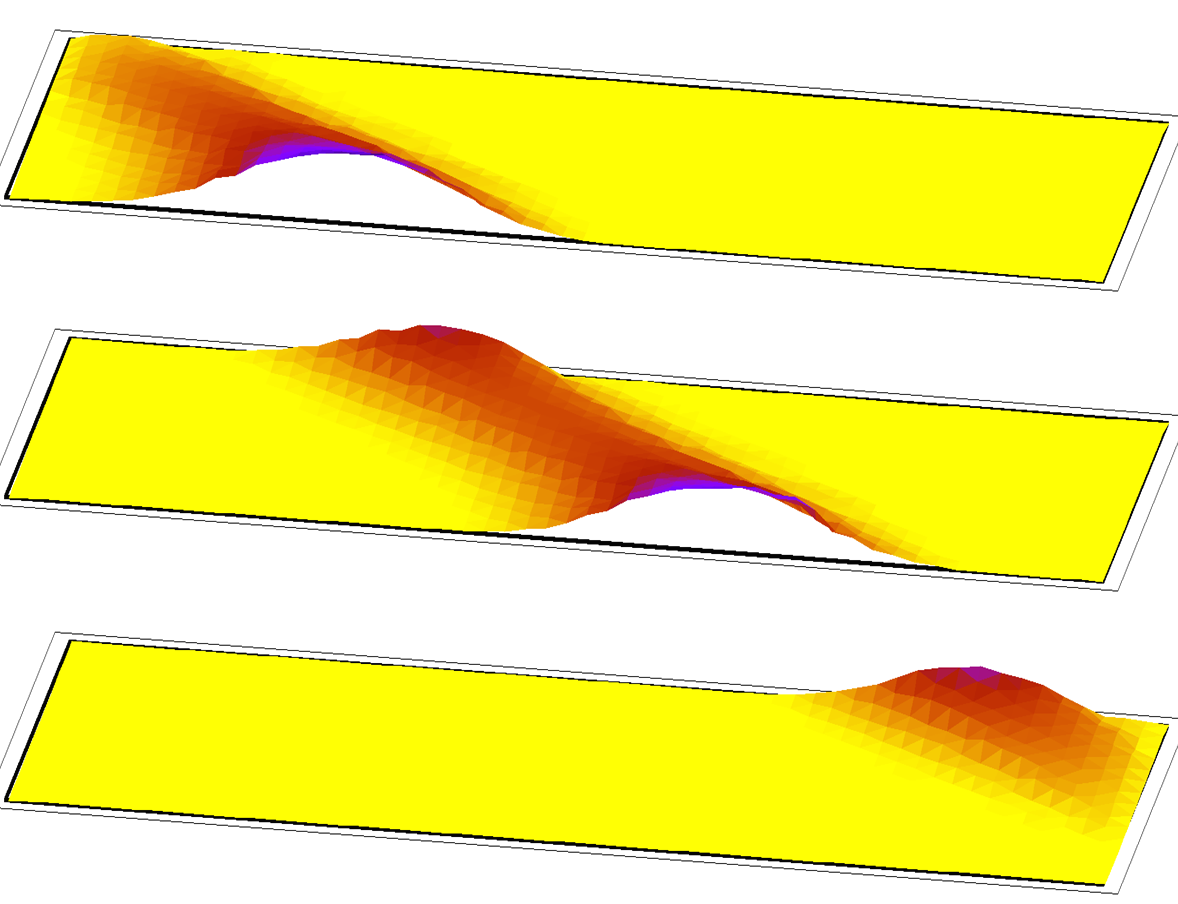}
& \includegraphics[width=.24\textwidth]{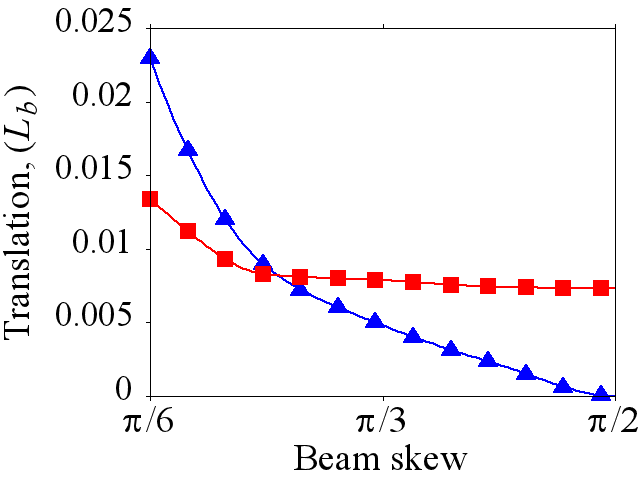} 
\end{tabular}
\caption{(a)  Successive shapes and positions of  a translating strip under a beam inclined by the angle $\pi/4$ to the director. (b) Dependence of velocities along (squares) and across (triangles) the director on the beam inclination to the director. 
Parameters: $\lambda=1.01$, $h=0.035$, $\eta=0.1$, $G=10^{-2}$.}
\label{f:skewed}
\end{figure}

\begin{figure}[b]
\begin{tabular}{cc} (a)& (b)\\
\includegraphics[width=.20\textwidth]{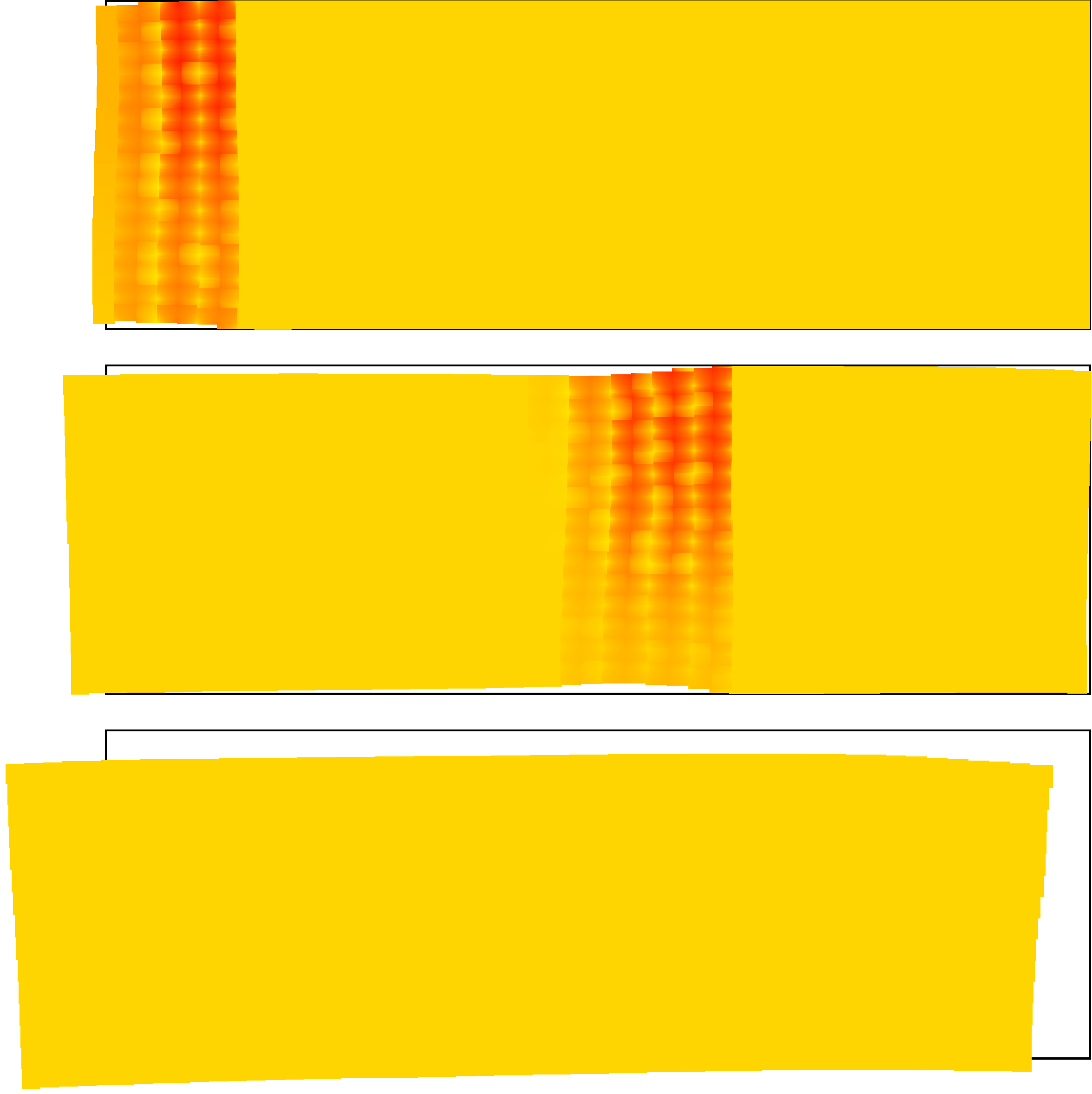} 
& \includegraphics[width=.26\textwidth]{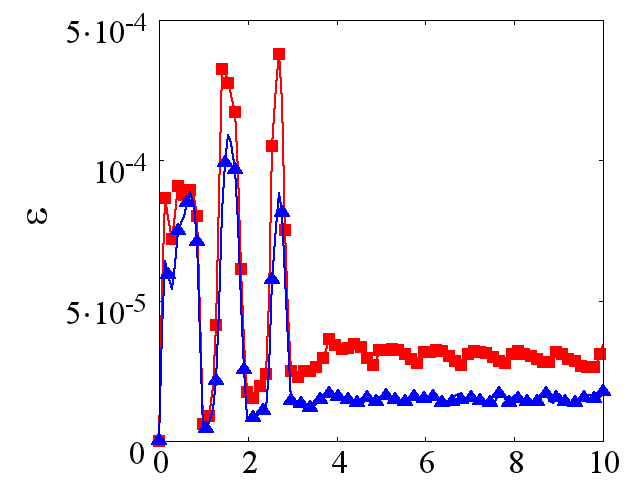} \end{tabular}
\caption{ (a) Snapshots of a stripe moving on the border of substrates with different surface mobilities during the 1st, 3rd, and 7th passage of the beam. Frames show the original position. Take note of buckling and/or narrowing of the stripe at the beam location.
 (b) Average energy along the edges in contact with high friction (squares) and low friction (triangles) substrates \emph{vs.} time (measured in the number of beam passages). Parameters: $h=0.05,\, \eta_1=0.1, \, \eta_2=0.01$.}
\label{f:fric}
\end{figure}

A sheet with the splayed director distribution compressed at the substrate side moves, on the opposite, along the beam propagation, as curving under the beam causes the projected length to decrease locally. With a splayed director, a moving strip, seen in Fig.~\ref{f-string}c, always bends, resting on its front and hind ``feet" only. Bending under the beam inducing NIT cannot be avoided, unless suppressed due to high flexural rigidity in thick sheets. The critical bending thickness, shown by the line marked by squares in Fig.~\ref{f-string}e, is constant everywhere except at very low surface mobility when bending is prevented by an extreme effort needed for shifting the front of the strip.  There is no motion under conditions that make bending impossible, both at large friction ($\eta \to 0$) and high flexibility ($h \to 0$) (to the left and below the lines marked by triangles in Fig.~\ref{f-string}e), and above the critical bending thickness. As a result, motion is suppressed below some critical surface mobility in this case, whereas this limit goes down to zero when the director is uniform. The propagation velocity reaches a maximum as a function of $h$ but increases with increasing mobility (beyond the right boundary of the plot), as back-sliding is avoided.  

More complex motion is observed in a strip actuated by an inclined beam (Fig.~\ref{f:skewed}). In this case, the stripe always bends, so that pure crawling is never observed. A skewed beam causes both the direction and velocity of motion to change, as shown in Fig.~\ref{f:skewed}; therefore the stripe may be guided along a curvilinear trajectory by changing beam inclination relative to the director. An increased velocity with the inclination angle $\psi$ (Fig.~\ref{f:skewed}b) is attributed to increasing activated area at a constant  beam length.

If a stripe is placed on the border of surfaces with different surface friction, it moves aside in the direction of higher mobility (Fig~\ref{f:fric}a). This can be evidently attributed to the tendency to energy minimization, as the stripe buckles stronger on the high-friction side thereby increasing its energy. As seen in Fig~\ref{f:fric}a, buckling, originally induced on the entire width of the stripe under the beam, fades away after a few beam passages, as the stripe shifts aside, thereby causing a sharp drop of energy on both edges (Fig~\ref{f:fric}b). The friction dependence is more straightforward here than in models of crawling cells where an optimal adhesion strength may be chosen \cite{igor}. The substrate stiffness (assumed here to be very high) should also affect nemato-elastic crawlers. 

\emph{Acknowledgement} This research is supported by Israel Science Foundation  (grant 669/14).


\end{document}